\documentclass[sigconf]{acmart}
\usepackage{graphicx}
\usepackage{amsmath}
\usepackage{bm}
\usepackage{bbm}
\usepackage{xcolor}
\usepackage{caption}
\usepackage{mathtools,xparse}
\usepackage{multirow}
\usepackage{balance}
\usepackage{threeparttable}
\usepackage{booktabs}
\usepackage{dsfont}
\usepackage{subcaption}
\newcommand{\etal}{\emph{et al. }}
\copyrightyear{2021}
\acmYear{2021}
\setcopyright{rightsretained}
\acmConference[DRL4IR '21]{The 2nd Workshop on Deep Reinforcement Learning for Information Retrieval}{July 15, 2021}{Virtual Event}
\acmPrice{}\acmDOI{}\acmISBN{}

\begin{document}
\title{Balancing Accuracy and Fairness for Interactive Recommendation with Reinforcement Learning}
%
% \titlerunning{FairRec}
% \author{Weiwen Liu\inst{1} \and
% Feng Liu\inst{2}\and
% Ruiming Tang\inst{3}\thanks{Corresponding to: \email{tangruiming2015@163.com, gychen@link.cuhk.edu.hk}}\and
% Ben Liao  \and
% Guangyong Chen\inst{4}$^\star$\and
% Pheng Ann Heng\inst{1,4}}
%
\author{Weiwen Liu}
\email{wwliu@cse.cuhk.edu.hk}
\affiliation{%
  \institution{The Chinese University of Hong Kong}
  \city{Hong Kong}
  \country{China}
}

\author{Feng Liu}
\email{fengliu@stu.hit.edu.cn}
\affiliation{%
  \institution{Harbin Institute of Technology}
  \city{Shenzhen}
  \country{China}
}

\author{Ruiming Tang}
\email{tangruiming2015@163.com}
\affiliation{
  \city{Shenzhen}
  \country{China}
}

\author{Ben Liao}
\email{liao@hotmail.co.uk}
\affiliation{
  \city{Shenzhen}
  \country{China}
}

\author{Guangyong Chen}
\email{gychen@link.cuhk.edu.hk}
\affiliation{%
  \institution{Shenzhen Institutes of Advanced Technology, Chinese Academy of Sciences}
  \city{Shenzhen}
  \country{China}
}

\author{Pheng Ann Heng}
\email{pheng@cse.cuhk.edu.hk}
\affiliation{%
  \institution{The Chinese University of Hong Kong}
  \city{Hong Kong}
  \country{China}
}
% \authorrunning{W. Liu et al.}
% \institute{The Chinese University of Hong Kong, Hong Kong \\ \and
% Harbin Institute of Technology, China \\ \and
% \email{tangruiming2015@163.com}\\ \and
% Guangdong Provincial Key Laboratory of Computer Vision and Virtual Reality Technology, Shenzhen Institutes of Advanced Technology, Chinese Academy of Sciences, China \\
% \email{gychen@link.cuhk.edu.hk}}

% \institute{The Chinese University of Hong Kong, Hong Kong \\
% \email{\{wwliu, pheng\}@cse.cuhk.edu.hk}\\ \and
% Harbin Institute of Technology, China \\
% \email{fengliu@stu.hit.edu.cn}\\ \and
% \email{tangruiming2015@163.com}\\ \and
% \email{liao@hotmail.co.uk}\\ \and
% Guangdong Provincial Key Laboratory of Computer Vision and Virtual Reality Technology, Shenzhen Institutes of Advanced Technology, Chinese Academy of Sciences, China \\
% \email{gychen@link.cuhk.edu.hk}}

% \institute{The Chinese University of Hong Kong, Hong Kong \\ \and
% Harbin Institute of Technology, China\\ \and
% Guangdong Provincial Key Laboratory of Computer Vision and Virtual Reality Technology, Shenzhen Institutes of Advanced Technology, Chinese Academy of Sciences, China}
%
\begin{abstract}
Fairness in recommendation has attracted increasing attention due to bias and discrimination possibly caused by traditional recommenders. In Interactive Recommender Systems (IRS), user preferences and the system's fairness status are constantly changing over time. Existing fairness-aware recommenders mainly consider fairness in static settings. Directly applying existing methods to IRS will result in poor recommendation. To resolve this problem, we propose a reinforcement learning based framework, \textit{FairRec}, to dynamically maintain a long-term balance between accuracy and fairness in IRS. User preferences and the system's fairness status are jointly compressed into the state representation to generate recommendations. FairRec aims at maximizing our designed cumulative reward that combines accuracy and fairness. Extensive experiments validate that FairRec can improve fairness, while preserving good recommendation quality.
\end{abstract}


\maketitle              % typeset the header of the contribution

% \vspace{-0.6cm}

%, failing to effectively maintain a balance between accuracy and fairness for IRS.
%
%
%

\section{Introduction}

Interactive Recommender Systems (IRS) have been widely implemented in various fields, \emph{e.g.,} news, movies, and finance~\cite{steck2015interactive}. Different from the conventional recommendation settings \cite{koren2009matrix}, IRS consecutively recommend items to individual users and receive their feedback in interactive processes. IRS gradually refine the recommendation policy according to the obtained user feedback in an online manner. The goal of such a system is to maximize the total utility over the whole interaction period.  A typical utility of IRS is user acceptance of recommendations. Conversion Rate (CVR) is one of the most commonly used measures of recommendation acceptance, computing the ratio of users \emph{performing a system's desired activity} to users \emph{having viewed recommended items}. A desired activity could be downloading from App stores, or making loans for microlending.

However, optimizing CVR solely may result in fairness issues, one of which is the unfair allocation of desired activities, like clicks or downloads, over different demographic groups. Under such unfair circumstances, majority (over-representing) groups may dominate recommendations, thereby holding a higher proportion of opportunities and resources, while minority groups are largely under-represented or even totally ignored. A fair allocation is a critical objective in recommendation due to the following benefits:

\textbf{Legal.}
Recommendation in particular settings are explicitly mandated to guarantee fairness. In the setting of employment, education, housing, or public accommodation, a fair treatment with respect to race, color, religion, \emph{etc.}, is required by the anti-discrimination laws \cite{holmes2005anti}. For job recommendation, it is expected that jobs at minority-owned businesses are being recommended and applied at the same rate as jobs at white-owned businesses. In microlending, loan recommender systems must ensure borrowers of different races or regions have an equal chance of being recommended and funded.

\textbf{Financial.}
Under-representing for some groups leads to the abandonment of the system. For instance, video sharing platforms like YouTube involve viewers and creators. It is desirable to ensure each creator has a fair chance of being recommended and promoted. Otherwise, if the new creators do not get adequate exposure and appreciation, they tend to leave the platform, resulting in less user-generated content. Consequently, users' satisfaction from both viewers and creators, as well as the platform's total income are affected in the long run.

The fairness concern in recommender systems is quite challenging, as accuracy and fairness are usually conflicting goals to be achieved to some extent. On the one hand, to obtain the ideal fairness, one could simply divide the recommendation opportunities equally to each item group, but users' satisfaction will be affected by being persistently presented with unattractive items. On the other hand, existing recommender systems have been demonstrated to favor popular items \cite{celma2008hits}, resulting in extremely unbalanced recommendation results. Thus, our work aims to answer this question: \emph{Can we achieve a fairer recommendation while preserving or just sacrificing a little recommendation accuracy?}

Most prior works consider fairness for the conventional recommender systems \cite{abdollahpouri2019multi,abdollahpouri2019beyond}, where the recommendation is regarded as a static process at a certain time instant. A general framework that formulates fairness constraints on rankings in terms of exposure allocation is proposed in \cite{singh2018fairness}. Individual attention fairness is discussed in \cite{biega2018equity}. \cite{surer2018multistakeholder} models re-ranking with fairness constraints in Multi-sided Recommender Systems (MRS) as an integer linear programming. The balanced neighborhoods method~\cite{burke2018balanced} balances protected and unprotected groups by reformulating the Sparse LInear Method (SLIM) with a new regularizer. 

However, it is hard to directly apply those methods to IRS due to: 

(i) It is infeasible to impose fairness constraints at every time instant. Forcing the system to be fair at any time and increasing fairness uniformly for all users will result in poor recommendations. In fact, IRS focus on the long-term cumulative utility over the whole interaction session, where the system could focus on improving accuracy for users with particular favor, and the lack of fairness at the time can later be compensated when recommending items to users with diversified interests. As such, we can achieve long-term system's fairness while preserving satisfying recommendation quality.

(ii) Existing work only considers the distribution of the number of recommendations (exposure) an item group received. Actually, the distribution of the desired activities that take place after an exposure like clicks or downloads has much larger commercial value and can be directly converted to revenue.

To resolve the above-mentioned problem, we design a novel \textbf{Fair}ness-aware \textbf{Rec}ommendation framework with reinforcement learning (FairRec) for IRS. FairRec jointly compresses the user preferences and the system's fairness status into the current state representation. A two-fold reward is designed to measure the system gain regarding accuracy and fairness. FairRec is trained to maximize the long-term cumulative reward to maintain an accuracy-fairness balance. The major contributions of this paper are as follows:
\begin{itemize}
    \item We formulate a fairness objective for IRS. To the best of our knowledge, this is the first work that balances accuracy and fairness in IRS.
    \item We propose a reinforcement learning based framework, FairRec, to dynamically maintain a balance between accuracy and fairness in IRS. In FairRec, user preferences and the system's fairness status are jointly compressed into the state representation to generate recommendations. We also design a two-fold reward to combine accuracy and fairness.
    \item We evaluate our proposed FairRec algorithm on both synthetic and real-world data. We show that FairRec can achieve a better balance between accuracy and fairness, compared to the state-of-the-art methods.
\end{itemize}

\begin{figure*}[t!]
  \centering
  \includegraphics[width=0.8\textwidth]{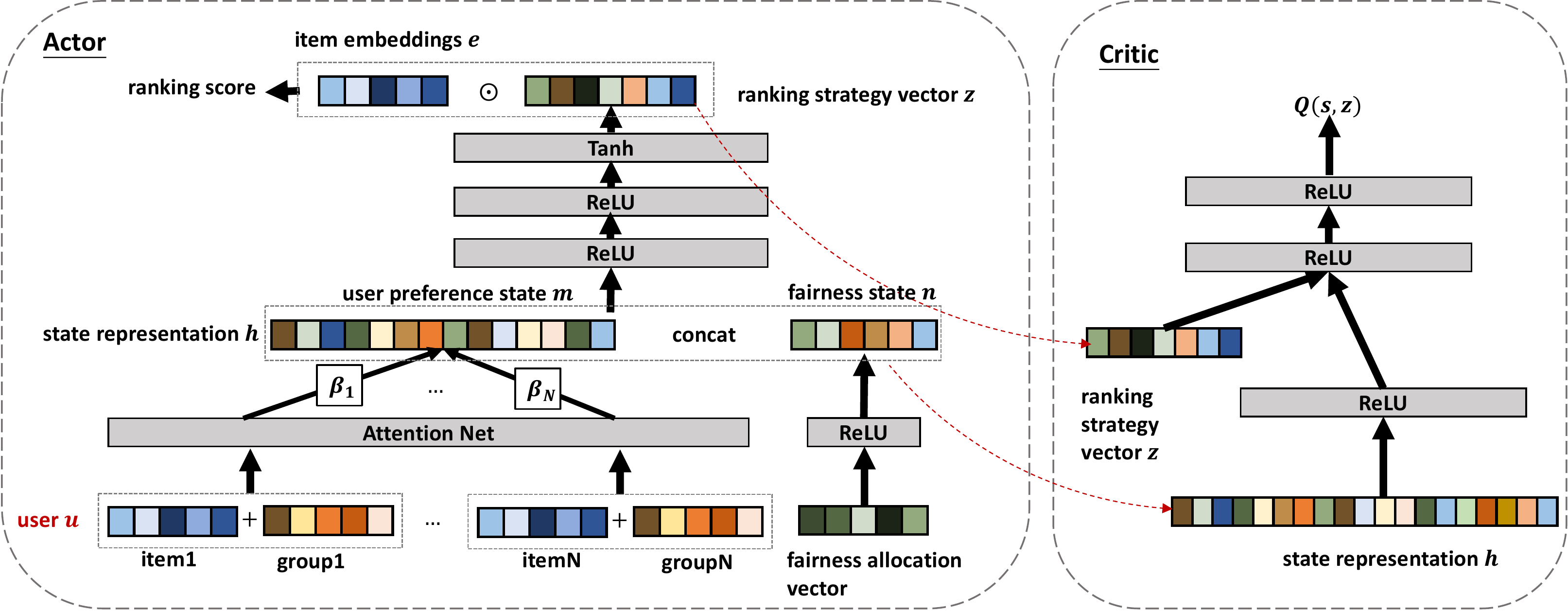}
  \caption{The architecture of FairRec.}\label{fig:fair_rl}
\end{figure*}

\section{Problem Formulation}\label{sec:problem_formulation}
\subsection{Markov Decision Process for IRS}
In this paper, we model the fairness-aware recommendation for IRS as a finite time Markov Decision Process (MDP), with an action space $\mathcal A$, a state space $\mathcal S$, and a reward function $r: \mathcal S\times \mathcal A\rightarrow \mathcal R$. When a user $u$ arrives at time step $t=1,\ldots,T$, the system observes the current state $s_t\in \mathcal S$ of the user $u$ and takes an action $a_t\in \mathcal A$ (\emph{e.g.}, recommending an item to the user). 

The user views the item and provides feedback $y_{a_t}$, \emph{e.g.}, clicking or downloading on the recommended item, if she feels interested. Let $y_{a_t}\in \{0, 1\}$ denote the user's feedback, with $y_{a_t}=1$ meaning the user performs desired activities, and $0$ otherwise. The system then receives a reward $r_t$ (a function of $y_{a_t}$), and updates the model. The problem formulation is formally presented as follows:

\textbf{States $\mathcal S$:} The state $s_t$ is described by user preferences and the system's fairness status. We jointly embed them into the current state representation. The detailed design of the state representation is given in Section \ref{subsec:state_representation}.

\textbf{Transitions $\mathcal P$:} The transition of states models the dynamic change of user preferences and the system's fairness. The successor state $s_{t+1}$ is obtained once the user's feedback at time $t$ is collected.

\textbf{Action $\mathcal A$:} An action $a_t$ is recommending an item chosen from the available candidate item set $\mathcal A$. Our framework can be easily extended to the case of recommending a list of items. To simplify our presentations, we focus on recommending an item at a time in this paper.

\textbf{Reward $\mathcal R$:} The reward $r_t$ is a scalar measuring the system's gain regarding accuracy and fairness after taking action $a_t$, elaborated in Section \ref{subsec:reward}.

We aim to learn a policy $\pi$, mapping from states to actions $a_t= \pi(s_t)$, to generate recommendations that are both accurate and fair. The goal is to maximize the sum of discounted rewards (return) from time $t$ onward, which is defined by $R_t^\gamma=\sum_{k=t}^T\gamma^{k-t}r_k$, and $\gamma$ is the discount factor.

\subsection{Weighted Proportional Fairness for IRS}
Each item is associated with a categorical protected attribute $C\in\{c_1,\ldots,c_l\}$. Let $\mathcal A_c=\{a|C=c, a\in \mathcal A\}$ denote the group of items with an attribute value $c$. Take loan recommendation for instance, if the protected attribute is the geographical region, then $\mathcal A_c$ with $c=\text{``Oceania"}$ contains all the loans applied from Oceania. Denote by $x_t\in \mathbb R^{l}_+$ the \emph{allocation vector}, where $x_t^i$ represents the allocation proportion of group $i$ up to time $t$,
\begin{equation}
x_t^i = \frac{\sum_{k=1}^t y_{a_k}\mathds{1}_{\mathcal A_{c_i}}(a_k)}{\sum_{i^{'}=1}^l\sum_{k=1}^t y_{a_k}\mathds{1}_{\mathcal A_{c_{i^{'}}}}(a_k)},\label{eq:x}
\end{equation}
where $\mathds{1}_{A}(x)$ equals to 1 if $x\in A$, and 0 otherwise. Recall that $y_{a_{k}}$ is the user's feedback on recommended item $a_k$. In loan recommendation, $x_t^i$ denotes the rate of funded loans from the region $i$ over all funded ones up to time $t$.

In this work, we focus on a well-accepted and axiomatically justified metric of fairness, the weighted proportional fairness \cite{kelly1998rate}. Weighted proportional fairness is a generalized Nash solution for multiple groups. 
\begin{definition}[Weighted Proportional Fairness]
An allocation of desired activities $x_t$ is weighted proportionally fair if it is the solution of the following optimization problem,
\begin{equation}\label{eq:fairness}
    \max_{x_t}\,\, \sum_{i=1}^lw_i\log(x^i_t), \quad\text{s.t.}\,\,\sum_{i=1}^lx^i_t = 1,\, x^i_t\geq0,\,i=1,\ldots,l.
\end{equation}
\end{definition}
The coefficient $w_i\in\mathbb R_+$ is a pre-defined parameter weighing the importance of each group. The optimal solution can be easily solved by standard Lagrangian multiplier methods, namely 
\begin{equation}
x_*^i=\frac{w_i}{\sum_{i^{'}=1}^lw_{i^{'}}}.
\end{equation}

As such, we aim to improve the weighted proportional fairness $\sum_{i=1}^lw_i\log(x_T^i)$ while preserving high conversions $\sum_{t=1}^Ty_{a_t}$ up to time $T$.

%: The actor takes as input of user's positively interacted items and the current allocation vector, and outputs a ranking strategy vector. The critic evaluates the ranking vector regarding the state representation. The agent selects the item with the largest ranking score, which is computed by the inner product of the item embeddings and the ranking strategy vector, with $\odot$ denotes the inner product.

\section{Proposed Model}
This section begins with a brief overview of our proposed FairRec. After that, we introduce the components of FairRec and the learning algorithm in detail.
\subsection{Overview}\label{subsection:proposed_model_overview}

To balance accuracy and fairness in the long run, we formulate IRS recommendation as an MDP, which is then solved by reinforcement learning.

The previously studied reinforcement learning models can be categorized as follows: \emph{Value-based} methods approximate the value function, then the action with the largest value is selected \cite{zheng2018drn,zhao2018recommendations}. Value-based methods are more sample-efficient and steady, but the computational cost is high when the action space is large. \emph{Policy-based} methods directly learn a policy that takes as input of the current user state and outputs an action \cite{chen2019large,wang2018reinforcement}, which generally have a faster convergence. \emph{Actor-critic} architectures take advantage of both value-based and policy-based methods \cite{zhao2018deep,liu2018deep}. Therefore, we design our model following the actor-critic framework.

The overall architecture of FairRec is illustrated in Figure \ref{fig:fair_rl}, which consists of an \emph{actor} network and a \emph{critic} network. The actor network performs time-varying recommendations according to the dynamic user preferences and the fairness status. The critic network estimates the value of the outputs associated with the actor network to encourage or discourage the recommended items. 

%We first introduce how to extract features from the current state $s_t$ in Section \ref{subsec:state_representation}, and then carefully design a reward function combining accuracy and fairness in Section \ref{subsec:reward}. Finally, the model update procedure is given in Section \ref{subsect: model_update}.

\subsection{Personalized Fairness-aware State Representation}\label{subsec:state_representation}
We propose a personalized fairness-aware state representation to jointly consider accuracy and fairness, which is composed of the the User Preference State (UPS) and the Fairness State (FS). State representation learns a non-linear transformation $h_t=f_s(s_t)$ that maps the current state $s_t$ to a continuous vector $h_t$. 

\smallskip \noindent
\textbf{User Preference State (UPS).}
UPS represents personalized user preferences. We propose a two-level granularity representation: the item-level and the group-level. The item-level representation indicates the user's fine-grained preferences to each item, while the group-level representation shows the user's coarse-grained interests in each item group. Such two-level granularity representation provides more information on the propensity of different users towards diverse recommendation. Therefore, the agent could focus on accuracy for the users with particular favor, and the lack of fairness at a point in time can later be compensated when recommending items to users with diverse interests.

The input of UPS is the sequence of the user $u$'s $N$ most recent positively interacted items, as well as the corresponding group IDs that the items belong to at $t$. Items belonging to the same group share the same protected attribute value $c$. Each item $a$ is mapped to a continuous embedding vector $e_a \in\mathbb R^d$. The embedding vector of each group ID $e_g$ is the average of the embedding vectors of all items belonging to the group $g$. Then each item is represented by
\begin{equation}
 \epsilon_{a} = e_a + e_g,
\end{equation}
where $\epsilon_{a}\in \mathbb R^d$, and item $a$ belongs to group $g$. The group embedding $e_g$ is added to serve as a global bias (or a regularizer), allowing items belonging to the same group to share the same group information.

As for a specific user $u$, the affects of different historical interactions on her future interest may vary significantly. To capture this sequential dependencies among the historical interacted items, we apply an attention mechanism \cite{vaswani2017attention} to weigh each item in the interacted item sequence. The attention net learns a weight vector $\beta$ of size~$N$, $\beta = \text{Softmax}(\omega^1\sigma(\omega^2[\epsilon_{a_1},\ldots,\epsilon_{a_N}]+b^2)+b^1)$, where $\omega^1,b^1,\omega^2,b^2$ are the network parameters and $\sigma(\cdot)$ is the ReLU activation function. The user preference state representation $m_t$ is obtained by multiplying the attention weights with the corresponding item representations as $ m_t = [\beta_1 \epsilon_{a_1},\ldots,\beta_N \epsilon_{a_N}]$,
% \vspace{-0.1cm}
% \begin{equation}
% m_t = [\beta_1 \epsilon_{a_1},\ldots,\beta_N \epsilon_{a_N}], 
% \vspace{-0.1cm}
% \end{equation}
where $m_t$ is of dimension $N\times d$ and $\beta_i$ denotes the $i$-th entry in the weight vector $\beta$. Therefore, the items currently contributing more to the outcome are assigned with higher weights.

\smallskip \noindent
\textbf{Fairness State (FS)}
The input of FS is the current allocation distribution of the desired activities at time $t$. As a complementary for UPS, FS provides evidence of the current fairness status and helps the agent to promote items belonging to under-represented groups. In particular, we deploy a Multi-Layer Perceptron (MLP) to map the allocation vector $x_t$ to a latent space, $n_t = \text{MLP}(x_t)$.
% \vspace{-0.1cm}
% \begin{equation}
% n_t = \text{MLP}(x_t). 
% \vspace{-0.1cm}
% \end{equation}
Then we concatenate $m_t$ and $n_t$ to obtain the final state representation, 
\begin{equation}
h_t = [m_t||n_t], 
\end{equation}
with $||$ denotes concatenation operation.

\subsection{Reward Function Design}\label{subsec:reward}
The reward is designed to measure the system's gain regarding accuracy and fairness. Existing reinforcement learning frameworks for recommendation only consider the recommendation accuracy, and one commonly used definition of reward is $r=1$ if the user performs desired activities and $-1$ otherwise \cite{liu2018deep,zhao2018deep}. To incorporate the fairness measure into IRS, we propose a two-fold reward by first examining whether the user performs the desired activities on the recommended item, and then evaluating the fairness gain of performing such a desired activity. 

As discussed in Section \ref{sec:problem_formulation}, to achieve the weighted proportional fairness, the optimal allocation vector is $x_*^i=\frac{w_i}{\sum_{i^{'}=1}^lw_{i^{'}}}$, with $w_i$ the pre-defined target allocation proportion of group $i$. Therefore, we incorporate the deviation from the optimal solution $x_*^i-x_t^i$ into the reward as the fairness indicator:
\begin{align}
r_t =
\begin{cases} \sum_{i=1}^l\mathds{1}_{\mathcal A_{c_i}}(a_t)\left(x_*^i-x_t^i+1\right), &\mbox{if } y_{a_t} = 1 \\
-\lambda, & \mbox{if } y_{a_t} = 0
\end{cases},
\end{align}
where $\mathds{1}_A(x)$ is the indicator function and is 1 when $x\in A$, 0 otherwise, $x_t^i$ is the allocation proportion of group $i$ at time $t$. The constant $\lambda>1$ is the penalty value for inaccurate recommendations and manages the accuracy-fairness tradeoff. A larger $\lambda$ means that the agent focuses more on accuracy. 

Since the fairness metric (Eq.~\eqref{eq:x} and Eq.~\eqref{eq:fairness}) is computed according to the number of the desired activities, only positive $y_{a_t}$ influences fairness. Therefore, we simply give a negative reward $-\lambda$ for $y_{a_t}=0$ to punish the undesired activities. When $y_{a_t}=1$, we compute the fairness score $x_*^i-x_t^i$, which is the difference between the optimal distribution and current allocation. Suppose the user performs a desired activity on the item $a_t\in \mathcal A_{c_i}$. Then the fairness score $x_*^i-x_t^i$ is negative if the $i$-th group is over-representing $(x_t^i>x_*^i)$, and is more negative if $\mathcal A_{c_i}$ already has a higher rate of the desired activity, indicating that the system should focus more on other groups. Similarly, the fairness score $x_*^i-x_t^i$ is positive if the $i$-th group is currently under-representing $(x_t^i<x_*^i)$, and is more positive if $\mathcal A_{c_i}$ is more lacking in the desired activity. We add 1 to the fairness score to ensure the reward is positive if $y_{a_t}=1$.

To sum up, the agent receives a large positive reward if the user performs a desired activity on the item and the item belongs to an under-representing group. Whereas the reward is a smaller positive number if the activity is desired, but the item belongs to an over-representing (majority) group. We punish the most severely with $y_{a_t}=0$, as it neither contributes to accuracy nor fairness.

\subsection{Model Update}\label{subsect: model_update}
\noindent
\textbf{Actor Network.}
The actor network extracts latent features from $s_t$ and outputs a ranking strategy vector $z_t$. The recommendation is performed according to the ranking vector by $a_t=\arg\max_{a\in\mathcal A}e_a^\top z_t$. In particular, we first embed $s_t$ to $h_t$ following the architecture described in Section \ref{subsec:state_representation}, then we stack fully-connected layers on top of $h_t$ to learn the nonlinear transformation and generate $z_t$, as presented in Figure \ref{fig:fair_rl}.

Suppose the policy $\pi_\theta(s)$ learned by the actor is parameterized by $\theta$. The actor is trained according to $Q_{\eta}(s_t,z_t)$ from the critic, and updated by the sampled policy gradient~\cite{silver2014deterministic} with $\alpha_\theta$ as the learning rate, $B$ as the batch size, 
\begin{equation}
\theta \leftarrow \theta + \alpha_\theta \frac{1}{B}\sum_t\nabla_z Q_\eta(s, z)|_{s=s_t,z=\pi_\theta(s_t)}\nabla_\theta\pi_\theta(s)|_{s=s_t},
\end{equation}

\noindent
\textbf{Critic Network.}
The critic adopts a deep neural network $Q_{\eta}(s_t, z_t)$, parameterized by $\eta$, to estimate the expected total discounted reward $\mathbb E[R_t^\gamma|s_t, z_t;\pi]$, given the state $s_t$ and the ranking strategy vector $z_t$ under the policy $\pi$. Specifically for this problem, the network structure is designed as follows
\begin{equation}
    Q_{\eta}(s_t, z_t) = \text{MLP}([\sigma(W_hh_t+b_h)||z_t]),
\end{equation}
by first mapping $h_t$ to the same space as $z_t$ with a fully-connected layer and then concatenating it with $z_t$, while $\text{MLP}(\cdot)$ denotes a mutli-layer perceptron, and $h_t=f_s(s_t)$ is the state representation as presented in Section \ref{subsec:state_representation}.

We use the temporal-difference (TD) learning~\cite{sutton1998introduction} to update the critic. The loss function is the mean square error $L=\sum_t(\nu_t-Q_\eta(s_t, z_t))^2$, where $\nu_t=r_t+\gamma Q_{\eta'}(s_{t+1},\pi_{\theta'}(s_{t+1}))$. The term $\nu_t-Q_\eta(s_t, z_t)$ is called time difference (TD), $\eta'$ and $\theta'$ are the parameters of the target critic and actor network that are periodically copied from $\eta, \theta$ and kept constant for a number of iterations to ensure the stability of the training \cite{lillicrap2015continuous}. The parameter $\theta$ is updated by gradient descent, with $\alpha_\eta$ the learning rate and $B$ the batch size:
\begin{equation}
    \eta\leftarrow \eta +\alpha_\eta \frac{1}{B}\sum_t(\nu_t-Q_\eta(s_t, z_t))\nabla_\eta Q_\eta(s_t, z_t).
\end{equation}

\begin{table*}[]
\centering
\small
\caption{Experimental results on MovieLens and Kiva.}\label{tab:results}
\begin{threeparttable}
\begin{tabular}{ccccccccc}
\toprule
                           &          & NMF             & SVD             & DeepFM          & LinUCB          & DRR             & MRPC            & FairRec         \\\midrule
\multirow{3}{*}{MovieLens} & CVR      & 0.7972 & 0.8478 & \underline{0.8612} & 0.8577 & 0.8592 & 0.8361 & \textbf{0.8702*}  \\
                           & PropFair & 0.8592 & 0.8337 & 0.8098 & 0.8464 & 0.8470 & \underline{0.8608} & \textbf{0.8666*}  \\
                           & UFG      & 4.2362 & 5.4795 & 5.8323 & 5.9476 & \underline{6.0177} & 5.2508 & \textbf{6.6776*}  \\\midrule
\multirow{3}{*}{Kiva}      & CVR      & 0.4211 & 0.4870 & 0.6349 & 0.6517 & \underline{0.6567} & 0.4286 & \textbf{0.6905*}  \\
                           & PropFair & 0.8473 & 0.8686 & 0.8671 & 0.8697 & 0.8645 & \underline{0.8761} & \textbf{0.8838*}  \\
                           & UFG      & 1.4635 & 1.6931 & 2.3752 & 2.4970 & \underline{2.5183} & 1.5332 & \textbf{2.8555*} \\
                           \bottomrule
\end{tabular}
\end{threeparttable}

\smallskip
\small
\raggedright   We conduct a two-sided significant test \cite{ruxton2006unequal} between FairRec and the strongest baseline DRR, where * means the p-value is smaller than 0.05.
\end{table*}

\section{Experiments}

\subsection{Experimental Settings}
We evaluate the proposed FairRec algorithm on both synthetic and real-world data, comparing with the state-of-the-art recommendation methods in terms of fairness and accuracy. 

\subsubsection{Datasets.}
We use MovieLens\footnote{https://grouplens.org/datasets/movielens} and Kiva.org datasets for evaluation. 

\textbf{MovieLens} is a public benchmark dataset for recommender systems, with 943 users, 1,602 items and 100,000 user-item interactions. Since the MovieLens data do not have protected attributes, we created 10 groups to represent differences among group inventories, and randomly assigned movies to each of such groups following a geometric distribution. An interaction with the rating (ranging from 1 to 5) larger than 3 is defined as a desired activity in calculating CVR.

\textbf{Kiva.org} is a proprietary dataset obtained from Kiva.org, consisting of lending transactions over a 6-month period. We followed the pre-processing technique used in \cite{liu2018personalizing} to densify the dataset. The retained dataset has 1,589 loans, 589 lenders and 43,976 ratings. The geographical region of loans is selected as the protected attribute, as Kiva.org has a stated mission of equalizing access to capital across different regions so that loans from each region have a fair chance to be funded. We define a transaction amount greater than USD25 as the desired activity for Kiva.

\subsubsection{Evaluation Metrics.}
We evaluate the recommendation accuracy by the Conversion Rate (CVR):
\begin{equation}
    \text{CVR} = \frac{\sum_{k=1}^Ty_{a_k}}{T},
\end{equation}
and measure the fairness by Weighted Proportional Fairness (PropFair)\footnote{The input of PropFair is shifted by one to avoid infinity results.}:
\begin{equation}
     \text{PropFair} =\sum_{i=1}^lw_i\log(1+x_T^i).
\end{equation}

Moreover, we propose a Unit Fairness Gain (UFG) to jointly consider accuracy and fairness,
\begin{equation}
\text{UFG}=\frac{\text{PropFair}}{\text{CVR}_{\text{max}}-\text{CVR}}=\frac{\text{PropFair}}{1-\text{CVR}}.
\end{equation}
UFG indicates the fairness of the system under unit accuracy budget. For any recommendation system, the ideal maximum CVR, namely $\text{CVR}_{\text{max}}$, equals to $1$. Thus UFG can be interpreted as the slope of fairness versus accuracy --- the fairness gain if we decrease a unit accuracy from $\text{CVR}_{\text{max}}$. A larger UFG means a higher value of PropFair can be achieved with unit deviation from CVR$_{\text{max}}$, namely, the larger, the better.

\subsubsection{Reproducibility.}
We randomly sample 80\% of the user with associated rating sequences for training, and 10\% for validation, 10\% for testing, so that the item dependencies within each session can be learned. We use grid search to select the hyper-parameters for all the methods to maximize the hybrid metric UFG: the embedding dimension in $\{10, 30, 50, 100\}$, the learning rate in $\{0.0001, 0.001, 0.01\}$. Embedding vectors are pre-trained using standard matrix factorization~\cite{koren2009matrix} following the traditional processing as in \cite{liu2018deep,zhao2018deep}. For the proposed FairRec, we set the number of recent interacted items  $N=5$, discount factor $\gamma=0.9$, the width of each hidden layer of the actor-critic network is 1000.  The batch size is set to 1024, and the optimization method is Adam. Without loss of generality, we set $w_i=1,i=1,\ldots,l$. All results are averaged from multiple independent runs.

\subsection{Results and Analysis}

\subsubsection{Comparison with Existing Methods.} We compare our proposed FairRec with six representative recommendation algorithms:
\begin{itemize}
    \item \textbf{NMF.} Non-negative Matrix Factorization (NMF) \cite{lee2001algorithms} estimates the rating matrix with positive user and item factors.
    \item \textbf{SVD.} Singular Value Decomposition (SVD) \cite{koren2010factor} is the classic matrix factorization based method that decomposes the rating matrix via a singular value decomposition.
    \item \textbf{DeepFM.} DeepFM \cite{guo2017deepfm} is the state-of-the-art deep learning model in recommendation that combines the factorization machines and deep neural networks.
    \item \textbf{LinUCB.} LinUCB \cite{li2010contextual} is the state-of-the-art contextual bandits algorithm that sequentially selects items and balances between exploitation and exploration in IRS.
    \item \textbf{DRR.} DRR \cite{liu2018deep} is a deep reinforcement learning framework designed for IRS to maximize the long-term reward.
    \item \textbf{MRPC.} Multi-sided Recommendation with Provider Constraints (MRPC) \cite{surer2018multistakeholder} is the state-of-the-art fairness-aware method by formulating the fairness problem as an integer programming.
\end{itemize}
% \textbf{(i) NMF.} Non-negative Matrix Factorization (NMF) \cite{lee2001algorithms} estimates the rating matrix with positive user and item factors; \textbf{(ii) SVD.} Singular Value Decomposition (SVD) \cite{DBLP:journals/tkdd/Koren10} is the classic matrix factorization based method that decomposes the rating matrix via a singular value decomposition; \textbf{(iii) DeepFM.} DeepFM \cite{guo2017deepfm} is the state-of-the-art deep learning model in recommendation that combines the factorization machines and deep neural networks; \textbf{(iv) LinUCB.} LinUCB \cite{li2010contextual} is the state-of-the-art contextual bandits algorithm that sequentially selects items and balances between exploitation and exploration in IRS; \textbf{(v) DRR.} DRR \cite{liu2018deep} is a deep reinforcement learning framework designed for IRS to maximize the long-term reward; \textbf{(vi) MRPC.} Multi-sided Recommendation with Provider Constraints (MRPC) \cite{surer2018multistakeholder} is the state-of-the-art fairness-aware method by formulating the fairness problem as an integer programming.

Table \ref{tab:results} shows the results. Bold numbers are the best results and underlined numbers are the strongest baselines. We have the following observations:

First, the deep learning based method (DeepFM) outperforms matrix factorization based methods (NMF and SVD) in CVR, while PropFair of DeepFM is lower. This is consistent with our expectation that DeepFM combines low-order and high-order feature interactions and has great fitting capability, yet it solely maximizes the accuracy, with fairness issues overlooked. 

Second, LinUCB and DRR generally achieve better CVR than matrix factorization and deep learning methods. It is because LinUCB and DRR consider the IRS setting, and aims to maximize the long-term reward. Compared LinUCB to DRR, LinUCB underperforms DRR since LinUCB assumes states of the system remain unchanged and fails to tailor the recommendation to match the dynamic user preferences. DRR is the strongest baseline as it achieves the best tradeoff between accuracy and fairness, with $\text{UFG}=6.0177$ on MovieLens and $\text{UFG}=2.5183$ on Kiva, respectively.

Third, MRPC considers fairness by adding fairness constraints for static recommendation. Therefore, MRPC generates the fairest recommendation on both datasets, but the CVR significantly decreases as MRPC ignores the dynamic change of user preferences and the fairness status.

Fourth, FairRec consistently yields the best performance in terms of CVR, PropFair, and UFG on both datasets, demonstrating FairRec is effective in maintaining the accuracy-fairness tradeoff over time. FairRec outperforms the strongest baselines, DRR, by 1.3\%, 2.3\%, and 11\% in CVR, PropFair, and UFG on MovieLens, and 5.1\%, 2.2\%, and 13.4\% on Kiva. Considering UFG, with unit accuracy loss, FairRec achieves the most fairness gain. FairRec observes the current user preferences and the fairness status, and estimates the long-term discounted cumulative reward. Therefore, FairRec is capable of long-term planning to manage the balance between accuracy and fairness.

\begin{figure}[t]
    \centering
    \hspace{1cm}
    \includegraphics[width=\linewidth]{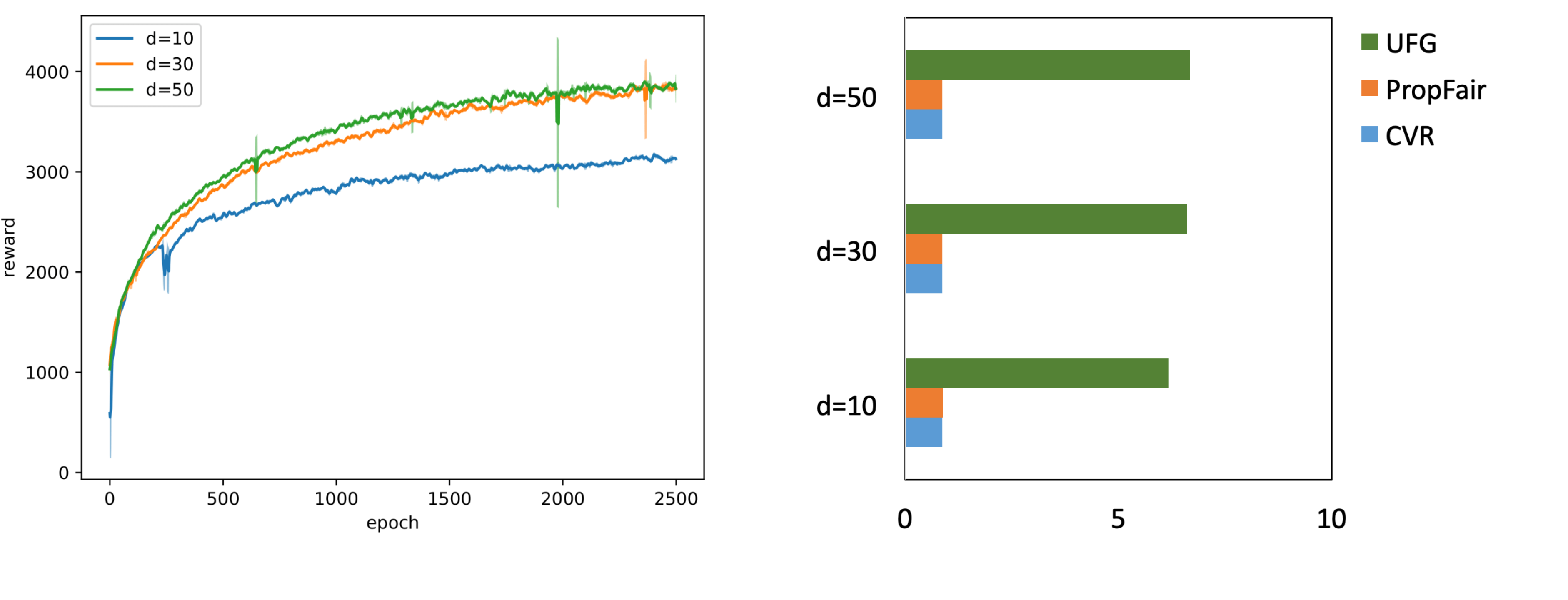}
    \caption{Experimental results with embedding dimension $d$ on MovieLens: cumulative reward (left) and CVR, PropFair, and UFG (right).}
    \label{fig:results}
\end{figure}

\subsubsection{Influence of Embedding Dimension.} Embedding dimension $d$ is an important factor for FairRec. We study how the embedding dimension $d$ influences the performance of FairRec. We vary $d$ in $\{10, 30, 50\}$, and run 2500 epochs. The cumulative reward and the test performance are plotted in Figure \ref{fig:results}.

We observe that when $d$ is large ($d=30$ and $d=50$), the algorithm benefits from sufficient expressive power and the reward converges at a high level. As for $d=10$, the cumulative reward converges fast at a relatively low value, indicating that the model suffers from the limited fitting capability. In terms of UFG value, $\text{UFG}=6.68$ when $d=50$, which is slightly better than 6.6 as $d=30$. Similar results can be found on Kiva, which is omitted for limited space. Therefore, we select $d=50$ in FairRec for all the experiments.

\subsubsection{Ablation Study.}
To evaluate the effectiveness of different components (\emph{i.e.}, the state representation and the reward function) in FairRec, we replace a component of FairRec with the standard setting in RL at each time, and compare the performance with the full-fledged FairRec. Experimental results are presented in Table~\ref{tab:ablation_study}. We design two variants: \textbf{FairRec(reward-)} with standard reward as in \cite{liu2018deep,zhao2018deep}; and \textbf{FairRec(state-)} with simple concatenation of item embeddings as the state representation as in \cite{liu2018deep}. 

Results show that FairRec(reward-) generally has high CVR, as no punishment on unfair recommendation. Moreover, the model simply optimizes accuracy, failing to balance accuracy and fairness. As for FairRec(state-), CVR is downgraded significantly, validating the importance of our designed state representation. Overall, UFG of FairRec is the largest, confirming that all the components of FairRec work together yield the best results.

\begin{table}[t]
\centering
\footnotesize
\caption{Ablation study on MovieLens and Kiva.}\label{tab:ablation_study}
\begin{threeparttable}
\begin{tabular}{ccccccc}
\toprule
                   & \multicolumn{3}{c}{MovieLens}                       & \multicolumn{3}{c}{Kiva}                            \\
                   & CVR             & PropFair        & UFG             & CVR             & PropFair        & UFG             \\\midrule
FairRec(reward-) & 0.8561          & 0.8053          & 5.5957          & \textbf{0.6935} & 0.8670          & 2.8290          \\
FairRec(state-)  & 0.8194          & \textbf{0.8758} & 4.8494          & 0.6723          & 0.8746          & 2.6688          \\\midrule
\textbf{FairRec}            & \textbf{0.8702} & 0.8666          & \textbf{6.6776} & 0.6905          & \textbf{0.8838} & \textbf{2.8555}  \\\bottomrule
\end{tabular}
\end{threeparttable}
\end{table}

\section{Related Work}
Our work is closely related to recommendation with deep reinforcement learning and fairness-aware recommendation.

\subsection{Recommendation with Reinforcement Learning}
Reinforcement Learning (RL) recommender systems recommend items by maximizing the long-term reward, where the reward can be the number of user repetitive clicks or purchases. Existing RL-based recommenders can be categorized into value-based methods \cite{zheng2018drn,zhao2018recommendations} and policy-based methods \cite{zhao2018deep,liu2018deep,hu2018reinforcement}.

Value-based methods compute Q-values of each item (or item subset) given a user state, and the one with the maximum Q-value is selected and recommended. Zheng \etal incorporate user return pattern as a supplement to click / no click feedback and adopt a Deep Q-learning framework for news recommendation \cite{zheng2018drn}. A DQN framework with Gated Recurrent Units (GRU) is used to capture users' positive and negative preferences simultaneously in \cite{zhao2018recommendations}. However, as value-based methods need to evaluate the Q-values of all the items under a specific user state \cite{wang2019privacy}, the computation becomes intractable when the number of items is large.

Policy-based methods directly learn a policy that takes as input of the current user state and outputs an action --- the item to be recommended. Zhao \etal utilized the Deep Deterministic Policy Gradient (DDPG) framework for page-wise recommendation \cite{zhao2018deep}. Various user state embeddings are studied in \cite{liu2018deep}. A deterministic policy gradient with full backup estimation (DPG-FBE) is proposed for learning the ranking function in \cite{hu2018reinforcement}. The ranking score of each item is further computed by the inner product of the item embeddings vectors and the learned ranking parameters.

\subsection{Fairness-aware Recommendation} 
Fairness and related concerns have become of increasing importance in recommender systems \cite{liu2019personalized}. Fairness is defined by the inverse of JS-Divergence between the actual exposure distribution and the desired exposure distribution and a post-processing method is proposed in \cite{modani2017fairness}. Rather than group fairness as we discussed in this chapter, individual fairness of each item is discussed in \cite{lee2014fairness}.  The balanced neighborhoods method~\cite{burke2018balanced} formulates the fairness problem into balancing protected and unprotected groups by imposing a regularizer on the Sparse Linear Method (SLIM). The fairness constraint is formulated as an integer programming optimization in \cite{surer2018multistakeholder}. However, all the existing methods (i) only consider the distribution of the number of recommendations (exposure) an item group received. Actually, the distribution of the desired actions like clicks or downloads is the major concern; (ii) perform the fairness-aware recommendation at every instant in time, leading to inferior recommendation results.

\section{Conclusion}
In this work, we propose a fairness-aware recommendation framework in IRS to dynamically balance accuracy and fairness in the long run with reinforcement learning. In the proposed state representation component, the user preference state (UPS) models both personalized preference and propensity to diversity; the fairness state (FS) is utilized to describe the current fairness status of IRS. A two-fold reward is designed to combine accuracy and fairness. Experimental results demonstrate the effectiveness in the balance of accuracy and fairness of our proposed framework over the state-of-the-art models.

\bibliographystyle{ACM-Reference-Format}
\bibliography{myref}

\end{document}